\documentclass[twocolumn,aps,prc,superscriptaddress,showpacs,floatfix]{revtex4}
\usepackage{graphicx}
\usepackage{epsfig}
\usepackage{pslatex}
\usepackage[normalem]{ulem}  
\usepackage[dvips]{color}
\renewcommand\sout{\bgroup \color{red} \ULdepth=-.5ex \ULset}

\begin{document}
\title{Thermal Charm Production in Quark-Gluon Plasma at LHC}

\author{Ben-Wei Zhang\footnote{On leave from Institute of Particle
Physics, Central China Normal University, Wuhan 430079, China}}
\author{Che Ming Ko}
\affiliation{Cyclotron Institute and Physics Department, Texas A\&M
University, College Station, Texas 77843-3366 USA}
\author{Wei Liu}
\affiliation{Cyclotron Institute and Physics Department, Texas A\&M
University, College Station, Texas 77843-3366 USA}


\begin{abstract}
Charm production from the quark-gluon plasma created in the
midrapidity of central heavy ion collisions at the Large Hadron
Collider (LHC) is studied in the next-to-leading order in QCD. Using
a schematic longitudinally boost-invariant and transversally
expanding fire-cylinder model, we find that charm production could
be appreciably enhanced at LHC as a result of the high temperature
that is expected to be reached in the produced quark-gluon plasma.
Sensitivities of our results to the number of charm quark pairs
produced from initial hard scattering, the initial thermalization
time and temperature of the quark-gluon plasma, and the charm quark
mass are also studied.
\end{abstract}



\maketitle

\section{Introduction}

In heavy-ion collisions at relativistic energies, hadrons composed
of confined quarks and gluons are expected to dissolve into their
constituents and form an extended volume of quark-gluon plasma
(QGP). Experiments at the Relativistic Heavy Ion Collider (RHIC)
have indeed shown that the results are consistent with the formation
of a strongly interacting quark-gluon plasma during the initial
stage of the collisions \cite{brahms,phobos,phenix,star}. Many
observables have been proposed to probe the properties of the
quark-gluon plasma. Among them is the production of particles
consisting of heavy charm and bottom quarks. In particular, it was
suggested that the production of charmonium $J/\psi$ in relativistic
heavy ion collisions might be suppressed as a result of its
dissociation in the produced quark-gluon plasma due to the Debye
screening \cite{matsui}. Although recent studies based on the
lattice QCD have indicated that the $J/\psi$ can survive in the
quark-gluon plasma at temperatures up to about twice the
deconfinement temperature \cite{asakawa,karsch}, $J/\psi$ production
in relativistic heavy ion collisions may still be suppressed in
heavy ion collisions at RHIC \cite{zhang00,Satz05,Rapp06}. However,
if the initial produced charm quark pairs are large such as in heavy
ion collisions at LHC, $J/\psi$ production could be enhanced through
regeneration from charm and anticharm quarks in the quark-gluon
plasma \cite{Thews06} as well as through statistical production from
\cite{stat03} or coalescence of \cite{greco04} charm and anticharm
quarks during hadronization of the quark-gluon plasma. These
mechanisms for $J/\psi$ production would become even more important
if there are other sources for charm quark production in
relativistic heavy ion collisions. To use charmonium and bottomonium
production in relativistic heavy ion collisions as a diagnostic tool
for the properties of produced quark-gluon plasma
\cite{Satz05,stat03,Thews06,Rapp06}, it is thus essential to
understand the production mechanism of charm and bottom quarks in
these collisions.

In heavy ion collisions, there are generally four different
contributions to charm production: direct and prethermal production
from partonic interactions as well as thermal production from
partonic and hadronic interactions. The direct production is from
initial hard scatterings of partons in the nucleons from the
colliding nuclei, which happens at a time scale of about $1/m_c \sim
0.15~{\rm fm}/c$ with $m_c$ being the charm quark mass. It
contributes about 3 charm quark pairs in one unit of midrapidity in
Au+Au collisions at $\sqrt{s_{NN}}=200~{\rm GeV}$ at RHIC, and this
number is increased to about 20 in Pb+Pb collisions at
$\sqrt{s_{NN}}=5.5~{\rm TeV}$ at LHC \cite{Vogt95,Vogt04}. These
numbers have, however, substantial uncertainty, particularly in
extrapolating to LHC energy from that at RHIC \cite{stat03,Vogt95},
where measurements from PHENIX for p+p collisions \cite{adare06} and
STAR for d+Au collisions \cite{adams05} disagree by about a factor
of 2. The prethermal production is from produced partonic matter
before it reaches thermal equilibrium. Based on consideration of
minijet partons, it has been shown in Ref.\cite{Lin95} that this
contribution is unimportant compared to that from direct production.
For thermal production, it can be from the thermalized partonic
matter or quark-gluon plasma as well as from the hadronic matter
formed after the hadronization of the quark-gluon plasma. For
thermal production of charm quark pairs from the quark-gluon plasma,
studies so far have been based on the lowest-order QCD process of
gluon-gluon fusion. The results show that its contribution depends
sensitively on the initial temperature of produced quark-gluon
plasma \cite{Matsui86,Biro90,Levai95,Levai97,Kampfer97}, and could
be important if the initial temperature is high. In hadronic matter,
charmed hadron can be produced from reactions such as $\pi N\to
D\Lambda_c$ \cite{cassing}, $\rho N\to D\Lambda_c$ \cite{liu02}, and
$NN\to N\Lambda_c D$ \cite{liu03}. According to Ref.\cite{hsd} based
on the Hadron-String Dynamics (HSD) without a partonic stage,
secondary meson-baryon reactions in the hadronic matter gives rise
to approximately 11\% enhancement of charm pair production in heavy
ion collisions at RHIC energies. This contribution is, however,
expected to be much smaller if one takes into account the formation
of the quark-gluon plasma in the collisions, which would lead to a
final hadronic matter with significantly lower density and
temperature than in the HSD model.

For Au+Au collisions at $\sqrt{s_{NN}}=200~{\rm GeV}$ at RHIC, charm
production is dominated by direct production as the contribution
from thermal production is negligible due to the relatively low
initial temperature ($\sim 350~{\rm MeV}$) of produced quark-gluon
plasma. Since the thermal production rate increases exponentially
with temperature as we will show later, thermal production of charm
quark pairs in Pb+Pb collisions at $\sqrt{s_{NN}}=5.5~{\rm TeV}$ at
LHC, in which a quark-gluon plasma with much higher initial
temperature and density than those at RHIC is expected to be
created, will be greatly enhanced. On the other hand, charm
production from initial hard scattering of colliding nucleons
increases only logarithmically with the collision energy. In most
studies on the production of charm particles at LHC, it has been
assumed, however, that thermal production can be neglected as well,
and only the direct production from initial hard scattering has been
considered. In the present study, we will consider thermal charm
production at LHC and compare its contribution to that due to direct
production to see whether this assumption can be justified or under
what conditions it can be considered as a good approximation.

Charm production in a quark-gluon plasma has been previously studied
in the lowest-order QCD \cite{Matsui86,
Biro90,Levai95,Levai97,Letessier07, Letessier-book}. In our study,
we will consider thermal charm production in the next-to-leading
order and investigate its sensitivity to the initial conditions of
the produced quark-gluon plasma at LHC. We find that based on a
reasonable estimation of the initial conditions, thermal charm
production could play an important role in determining the total
number of charm pairs produced in relativistic heavy ion collisions.

This paper is organized as follows. In the next section, we discuss
the processes and corresponding cross sections for thermal charm
production in the leading order as well as the next-to-leading order
in QCD. In Section III, we introduce the time evolution of formed
quark-gluon plasma in heavy ion collisions at LHC and derive the
rate equation for charm production in the quark-gluon plasma. In
Section IV, we present the numerical results on thermal charm
production at LHC and their dependence on the initial conditions and
other parameters in the model. Finally, a summary and conclusions
are given in Section V.

\section{charm production from quark and gluon interactions}

\subsection{charm production cross sections}

Previous studies of thermal charm production in a quark-gluon plasma
\cite{Matsui86,Biro90,Levai95,Levai97,Kampfer97} were all based on
the following leading-order processes in QCD:
\begin{eqnarray}\label{born}
q +\bar{q}&\rightarrow& c + \bar{c},\\
g + g&\rightarrow& c + \bar{c}.
\end{eqnarray}
In the present study, we include also the next-leading-order
processes
\begin{eqnarray}\label{radiative}
q + \bar{q}&\rightarrow& c + \bar{c} + g,\\
g + g&\rightarrow& c+\bar{c}+ g, \label{process-2gluon}
\end{eqnarray}
and the interferences between the leading-order processes with their
virtual corrections due to vertex corrections and self energy
insertions as their contributions are of same order in the QCD
coupling as the next-leading-order processes. We neglect, however,
next-leading-order processes involving the gluon-quark interactions
\begin{eqnarray}
g + q&\rightarrow& c + \bar{c} + q,\nonumber\\
g + \bar{q}&\rightarrow& c + \bar{c} + \bar{q}, \label{gluon+quark}
\end{eqnarray}
as their contributions are less important compared to other
processes, especially the gluon-gluon fusion process of
Eq.(\ref{process-2gluon}). This is largely due to the smaller color
factor from the gluon-quark interaction vertex as compared to that
from the gluon-gluon vertex, and the fact that the charm pair can be
produced from two incoming gluon lines in gluon-gluon fusion
process, whereas in the gluon-quark process it can only come from a
single gluon line \cite{Nason}.

\begin{figure}[h]
\vspace{0.45cm}
\centerline{\includegraphics[width=1.5in,height=2in]{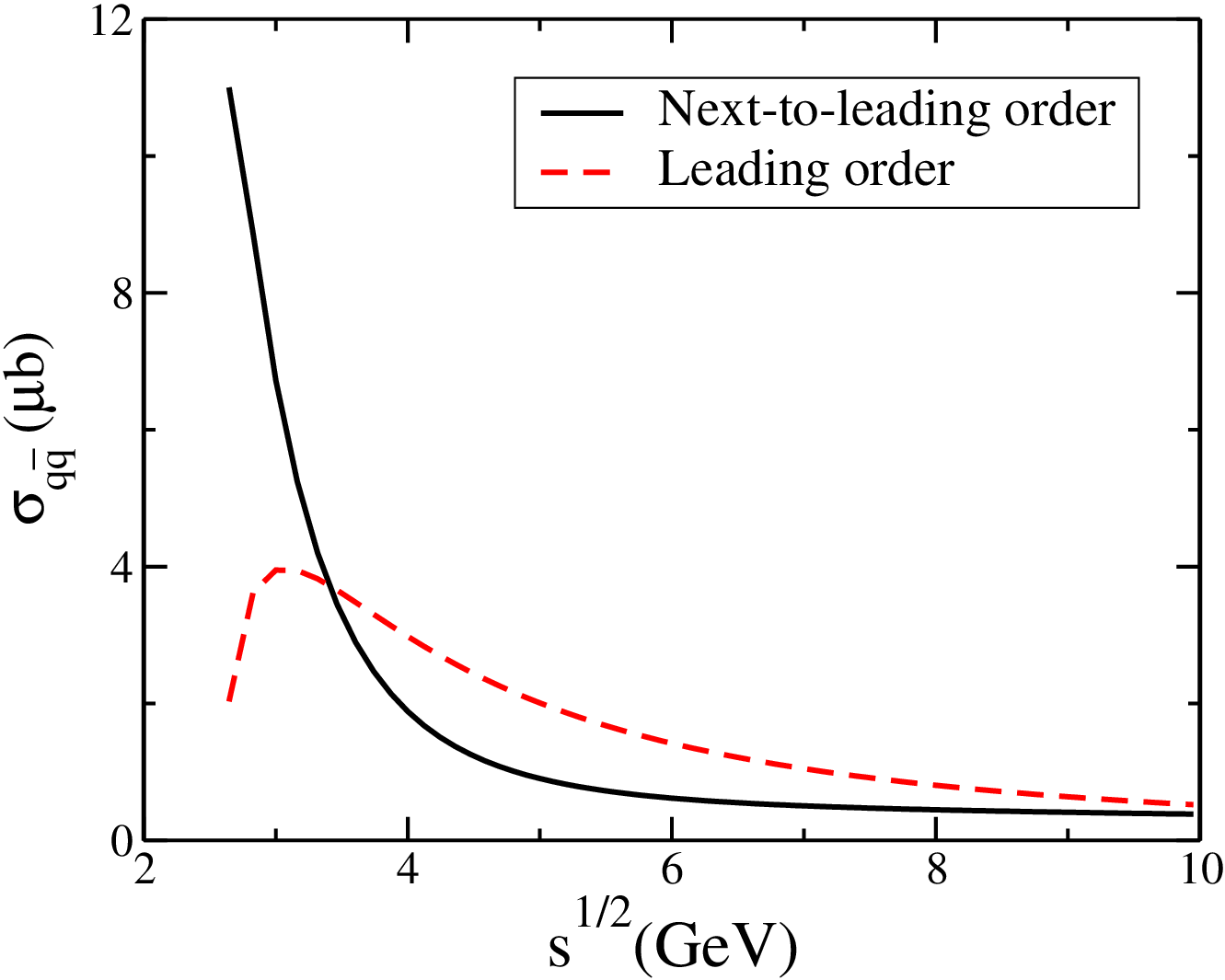}
\hspace{0.4cm}
\includegraphics[width=1.5in,height=2in]{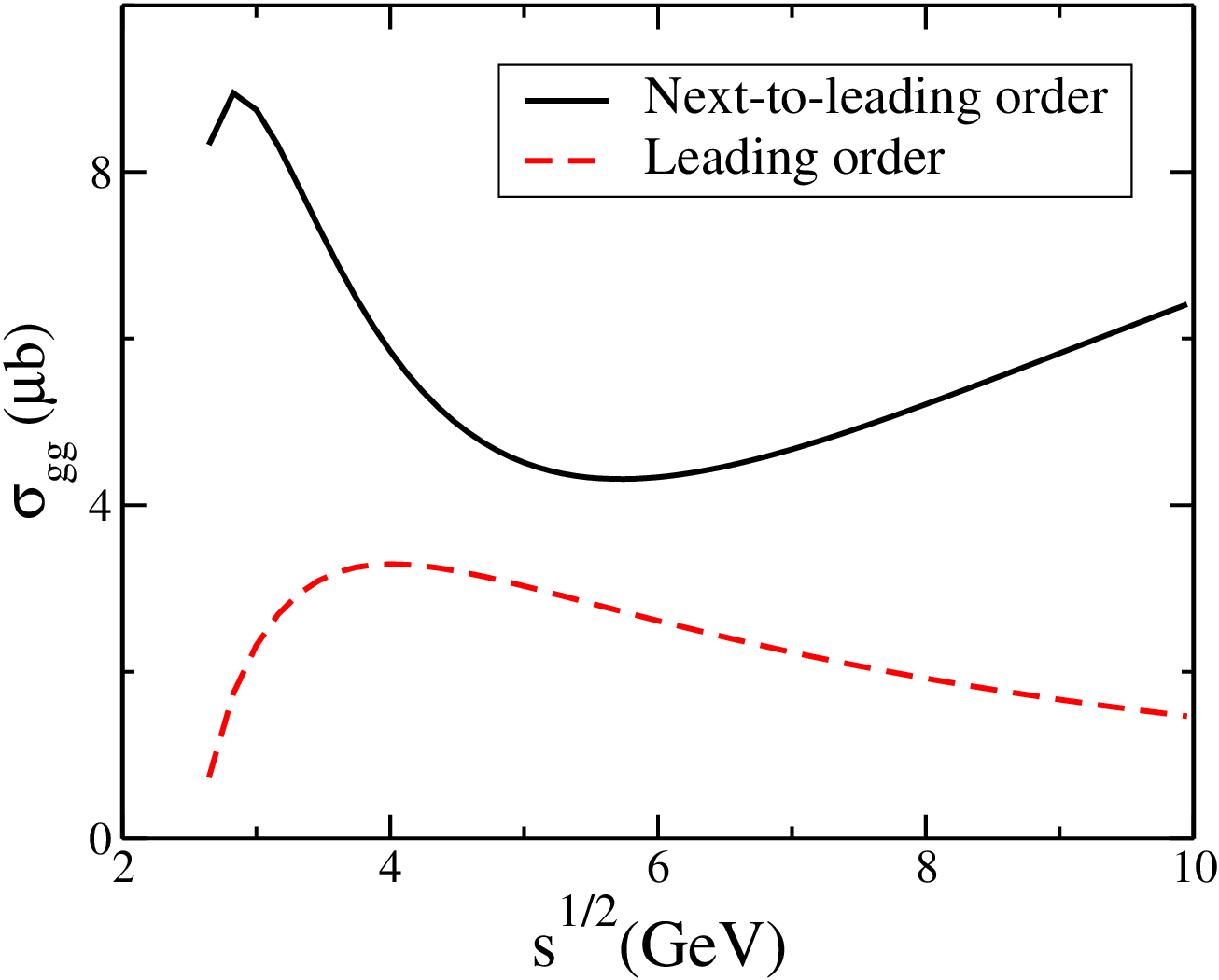}}
\caption{(Color online) Charm quark pair production cross sections
as functions of center-of-mass energy from quark-antiquark
annihilation (left panel) and gluon-gluon fusion (right panel) in
both the leading (dashed line) and the next-to-leading (solid line)
order. The charm quark mass is taken to be $m_c=1.3~{\rm GeV}$,
while light quarks and gluons are massless.} \label{sigma}
\end{figure}

The total cross section for charm production in the
next-to-leading-order in QCD has been previously derived. In the
present study, we use the results given in Refs.
\cite{Beenakker,Nason}, where the soft divergences in the
next-leading-order $2\rightarrow 3$ processes are eliminated by
corresponding divergences in virtual corrections to the
leading-order $2\to 2$ processes, while the ultraviolet divergences
in virtual corrections are handled by renormalization in the
modified minimal-subtraction ($\overline{MS}$) scheme
\cite{Field,Peskin}. Choosing the QCD running coupling constant as
$\alpha_s(m_c)$ and the renormalization scale as the charm quark
mass, we have calculated the cross sections for the production of
charm pairs with charm quark mass $m_c=1.3~{\rm GeV}$ from massless
light quarks and gluons, and they are shown in Fig.~\ref{sigma} as
functions center-of-mass energy for the quark-antiquark annihilation
(left panel) and gluon-gluon fusion (right panel) processes. The
dashed and solid lines correspond to results for the leading order
and the next-to-leading order, respectively. It is seen that for
charm production from quark-antiquark annihilation, the
next-to-leading order gives a larger cross section at low energies
but a smaller one at high energies. The cross section for charm
production from gluon-gluon fusion is, on the other hand,
significantly larger in the next-to-leading order than in the
leading order at all energies.

\subsection{thermal averaged charm production cross sections}

In the kinetic model to be used in the next section for studying
charm quark production and annihilation in a quark-gluon plasma,
thermal averaged cross sections are needed. In terms of the thermal
distribution functions $f_i(\mathbf{p})$ of quarks and gluons in the
quark-gluon plasma and the relative velocity $v_{ab}$ of two initial
interacting partons $a$ and $b$, the thermal averaged cross section
$\sigma _{ab\rightarrow cd}$ for the reaction $ab\to cd$ is given by
\cite{ko}
\begin{eqnarray}
\left\langle \sigma _{ab\rightarrow cd}v\right\rangle &=&\frac{\int
d^{3}\mathbf{p}_{a}d^{3}\mathbf{p}_{b}f_{a}(\mathbf{p}_{a})f_{b}(\mathbf{p}_{b})
\sigma _{ab\rightarrow cd}v_{ab}}{\int d^{3}\mathbf{p}_{a}d^{3}
\mathbf{p}_{b}f_{a}(\mathbf{p}_{a})f_{b}(\mathbf{p}_{b})}\nonumber\\
&=&[4\alpha^2_a K_2(\alpha_a) \alpha^2_b K_2(\alpha_b)]^{-1}  \nonumber \\
&\times&\int^\infty_{z_0} dz[z^2-(\alpha_a+\alpha_b)^2]
[z^2-(\alpha_a-\alpha_b)^2]\nonumber\\
&\times&K_1(z)\sigma(s=z^2 T^2),
\end{eqnarray}
with $\alpha_i=m_i/T$, $z_0=\mathrm{max}(\alpha_a+\alpha_b,\alpha_c
+\alpha_d)$, and $K_1$ being the modified Bessel function.

\begin{figure}[h]
\vspace{0.5cm} \centerline{
\includegraphics[width=1.5in,height=2in]{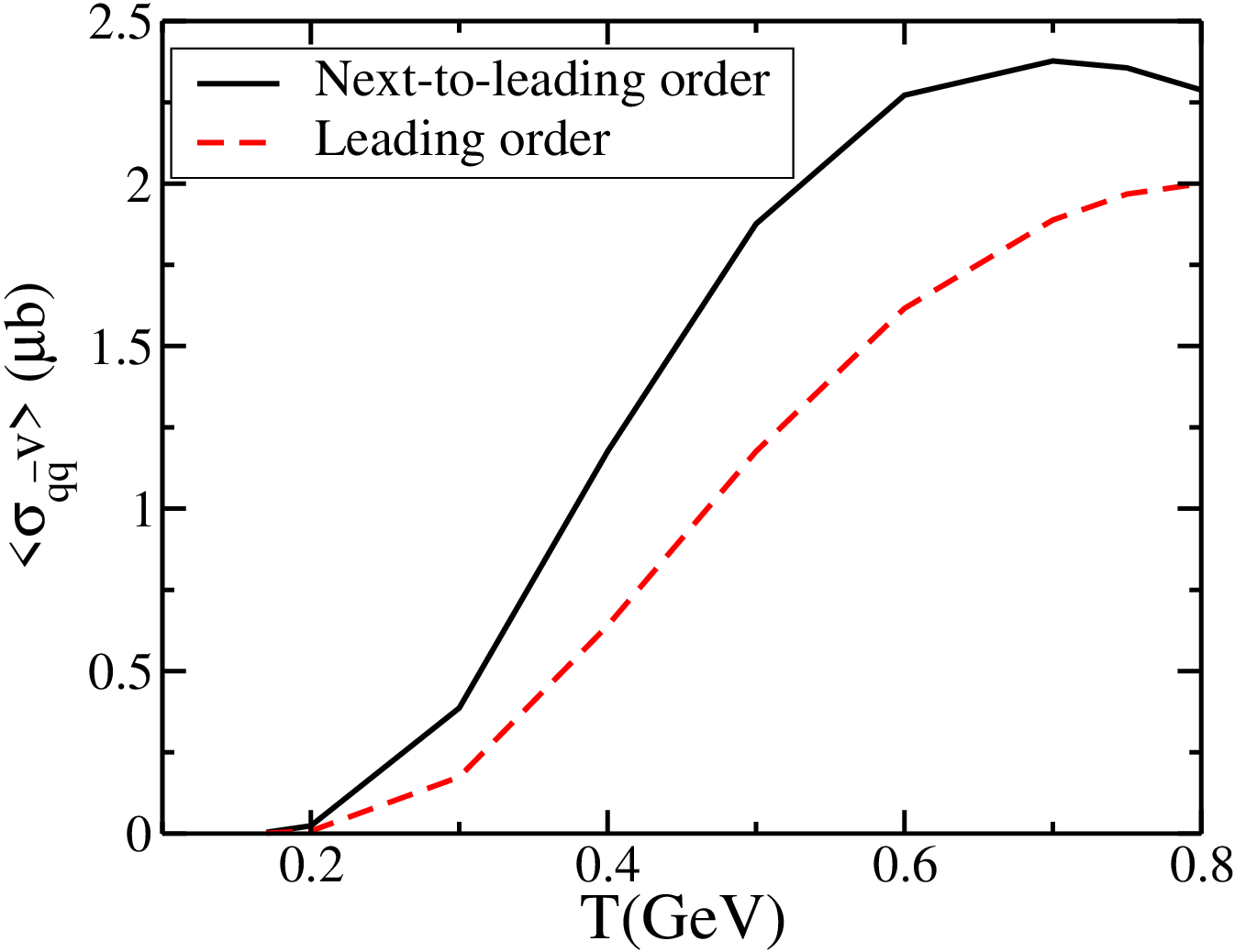}
\hspace{0.3cm}
\includegraphics[width=1.5in,height=2in]{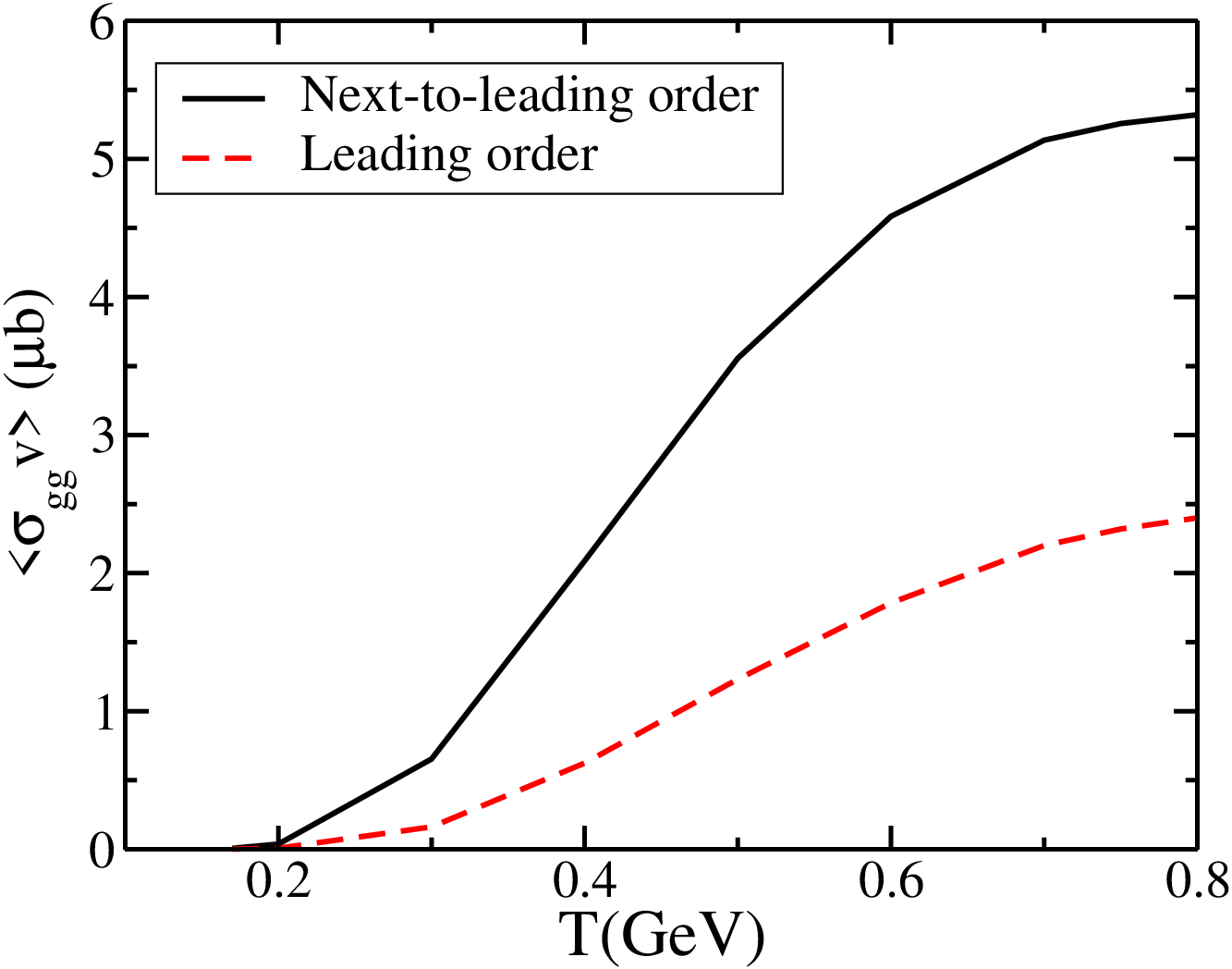}}
\caption{(Color online) Thermal averaged cross sections for charm
pair production from quark-antiquark annihilation (left panel) and
gluon-gluon fusion in a quark-gluon plasma as functions of
temperature. The charm quark mass is taken to be $m_c=1.3~{\rm GeV}$
while quarks and gluons are taken to have thermal masses given by
Eq.~(\ref{thermal}).} \label{sigmav}
\end{figure}

In the quark-gluon plasma, quarks and gluons acquire thermal masses.
As an exploratory study, we include this effect in their
distribution functions but not in the calculation of their
scattering cross sections. For the thermal masses of quarks and
gluons, they are taken to be \cite{bellac,blazoit}:
\begin{eqnarray}\label{thermal}
m_q = gT/\sqrt{6}~~{\rm and}~~m_g=gT/\sqrt{2},
\end{eqnarray}
where $g$ is the QCD coupling constant and is taken to have the
value $g=\sqrt{4\pi\alpha_s(2\pi T)}$. With the charm quark mass
$m_c=1.3~{\rm GeV}$, the thermal averaged cross section for charm
production in a quark-gluon plasma is shown in Fig.~\ref{sigmav} as
function\sout{s} of the temperature of the quark-gluon plasma for
quark-antiquark annihilation (left panel) and gluon-gluon fusion
(right panel). For both reactions, thermal averaged cross sections
are significantly larger in the next-to-leading order (solid line)
than in the leading order (dashed line).

\section{charm production in heavy ion collisions at LHC}

Using above thermal averaged charm production cross sections, we
study in this section the time evolution of the abundance of charm
quark pairs in heavy ion collisions at LHC using a kinetic model
based on the rate equation that takes into account both production
and annihilation of charm quarks in the produced quark-gluon plasma.
For the dynamics of the quark-gluon plasma, we describe it by a
schematic hydrodynamic model and assume that both quarks and gluons
are in thermal and chemical equilibrium during the evolution. We
further assume that all produced charm quarks including those
produced directly from the initial hard collisions are also in
thermal equilibrium, although not in chemical equilibrium. The
latter is consistent with observed large elliptic flow of the
electrons from charmed meson decays in heavy ion collisions at RHIC
\cite{adler,laue}, which requires that charm quarks interact
strongly in the quark-gluon plasma and thus are likely to reach
thermal equilibrium \cite{hees,zhang,molnar}.

\subsection{the rate equation}

The time evolution of the number density of charm quark pairs
$n_{c{\bar c}}$ in an expanding quark-gluon plasma can be described
by the rate equation \cite{Ko3,Koch02}:
\begin{eqnarray}
\partial_\mu(n_{c \bar c} u^\mu)&=&R_{q{\bar q}\to c\bar c}
+R_{q{\bar q}\to c\bar cg}+ R_{g g\to c\bar c}+R_{g g\to c\bar cg}\nonumber\\
&-&R_{c \bar c\rightarrow q {\bar q}}-R_{c \bar cg\rightarrow q
{\bar q}}-R_{c \bar c \rightarrow g g}-R_{c \bar cg \rightarrow g
g}. \label{rate-eq:1}
\end{eqnarray}
In the above, $u^\mu=\gamma(1,\mathbf{v})$ is the four velocity of a
fluid element in the quark-gluon plasma with velocity $\mathbf{v}$
and the corresponding Lorentz factor $\gamma$, and the terms on the
left hand side of above equation are the charm pair production and
annihilation rates. To the next-to-leading order, the charm pair
production rate is given by
\begin{eqnarray}
R_{q{\bar q}\rightarrow c \bar c}&=& \langle\sigma_{q{\bar
q}\rightarrow c \bar c}v\rangle n_{q} n_{{\bar q}},\nonumber\\
R_{g g\rightarrow c \bar c}&=& \frac{1}{2}\langle\sigma_{g g
\rightarrow c \bar c}v\rangle n_{g}^2, \nonumber\\
R_{q{\bar q}\rightarrow c \bar cg}&=& \langle\sigma_{q {\bar
q}\rightarrow c \bar c g}v\rangle n_{q} n_{{\bar q}}, \nonumber\\
R_{g g\rightarrow c \bar cg}&=& \frac{1}{2}\langle\sigma_{g g
\rightarrow c \bar c g}v\rangle n_{g}^2,
\end{eqnarray}
where $n_g$, $n_q$ and $n_{\bar q}$ denote the gluon, quark, and
antiquark densities in the quark-gluon plasma, respectively, and
they are taken to have their equilibrium values. The leading-order
cross sections $\sigma_{q\bar q\to c\bar c}$ and $\sigma_{g\bar g\to
c\bar c}$ in above equation are computed from the processes in
Eq.(\ref{born}), while the next-leading-order cross sections
$\sigma_{q\bar q\to c\bar cg}$ and $\sigma_{g\bar g\to c\bar cg}$
include both the processes in Eq.(\ref{radiative}) and the virtual
corrections to the leading-order processes.

\vspace{1.3cm}
\begin{figure}[ht]
\centerline{
\includegraphics[width=2.5in,height=2.5in,angle=0]{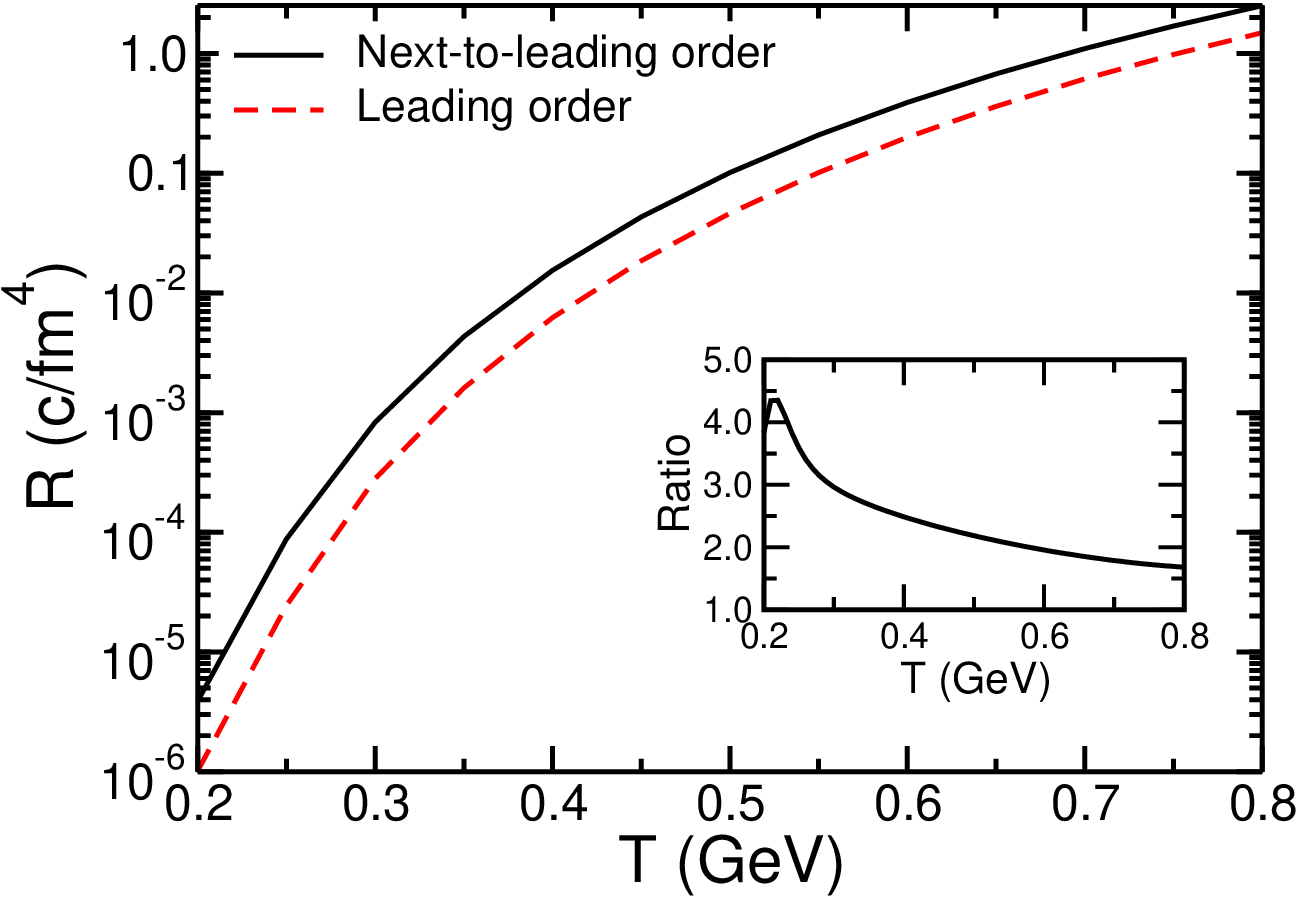}}
\caption{(Color online) Thermal charm quark production rate as a
function of temperature for massive partons and charm quark mass
$m_c=1.3~{\rm GeV}$. The inset gives the ratio of the charm
production rate in the next-to-leading order to that in the leading
order.} \label{rate}
\end{figure}

In Fig.~\ref{rate}, we show the thermal charm production rate as a
function of temperature with massive quarks and gluons in a
thermally and chemically equilibrated quark-gluon plasma. It is seen
that the charm production rate increases almost exponentially with
increasing temperature. With the initial temperature $T_0\simeq 350$
MeV in central heavy ion collisions at RHIC, the thermal charm
production rate is more than two orders of magnitude smaller than
that at the temperature $T_0\simeq 700$ MeV expected to be reached
in central heavy ion collisions at LHC. This thus justifies the
neglect of thermal charm production at RHIC. The inset in
Fig.~\ref{rate} gives the ratio of the charm production rate in the
next-to-leading order to that in the leading order, which is seen to
vary from $\sim 4.5$ at low temperatures to $\sim 1.8$ at high
temperatures.

Since the charm pair annihilation rate and production rate become
equal when the charm quark abundance reaches chemical equilibrium,
i.e,
\begin{eqnarray}
R^{\mathrm{eq}}_{c \bar c \rightarrow q{\bar q}} &=&
R^{\mathrm{eq}}_{q{\bar q} \rightarrow c \bar c} =
\langle\sigma_{q{\bar q}\rightarrow c \bar c}v\rangle n^{\rm eq}_{q}
n^{\mathrm{eq}}_{ {\bar q}},\nonumber\\
R^{\mathrm{eq}}_{c\bar c\rightarrow g g} &=& R^{\mathrm{eq}}_{g
g\rightarrow c \bar c} = \frac{1}{2}\langle\sigma_{g g \rightarrow c
\bar c}v\rangle (n^{\mathrm{eq}}_{g})^2,  \nonumber \\
R^{\mathrm{eq}}_{c \bar c g\rightarrow q{\bar q}} &=&
R^{\mathrm{eq}}_{q{\bar q} \rightarrow c \bar c g} =
\langle\sigma_{q{\bar q}\rightarrow c \bar c g}v\rangle
n^{\rm eq}_{q}n^{\mathrm{eq}}_{ {\bar q}},\nonumber\\
R^{\mathrm{eq}}_{c\bar cg\rightarrow g g} &=& R^{\mathrm{eq}}_{g g
\rightarrow c \bar c g} = \frac{1}{2}\langle\sigma_{g g \rightarrow
c \bar c g}v\rangle (n^{\mathrm{eq}}_{g})^2,
\end{eqnarray}
where $n^{\mathrm{eq}}$ is the equilibrium density while
$R^{\mathrm{eq}} $ is the rate evaluated with the equilibrium
densities, the charm pair annihilation rates in the leading order
and the next-leading order can thus be written as
\begin{eqnarray}
R_{c \bar c\rightarrow q{\bar q}} &=& \langle\sigma_{q {\bar q}
\rightarrow c \bar c}v\rangle (n^{\rm eq}_{q})^2\left(\frac{n_{c\bar
c}}{n^{\mathrm{eq}}_{c\bar
c}}\right)^2,\nonumber\\
R_{c \bar c\rightarrow g g} &=& \frac{1}{2} \langle\sigma_{g g
\rightarrow c \bar c}v\rangle (n^{\rm eq}_{g})^2
\left(\frac{n_{c\bar c}}{n^{\mathrm{eq}}_{c\bar c}}\right)^2,
\nonumber\\
R_{c \bar c g\rightarrow q{\bar q}} &=& \langle\sigma_{q
{\bar q} \rightarrow c \bar c g}v\rangle (n^{\rm
eq}_{q})^2\left(\frac{n_{c\bar c}}{n^{\mathrm{eq}}_{c\bar
c}}\right)^2,\nonumber\\
R_{c \bar c g\rightarrow g g} &=& \frac{1}{2} \langle\sigma_{g g
\rightarrow c \bar c g}v\rangle (n^{\rm eq}_{g})^2
\left(\frac{n_{c\bar c}}{n^{\mathrm{eq}}_{c\bar c}}\right)^2.
\end{eqnarray}
In the above, we have assumed that the density of quarks is the same
as that of antiquarks, i.e., $n_q^{\rm eq}=n_{\bar q}^{\rm eq}$.

With above equations and taking light quarks and gluons to be in
thermal and chemical equilibrium, the rate equation
Eq.~(\ref{rate-eq:1}) can be rewritten as
\begin{eqnarray}\label{rate-eq:2}
&&\partial_\mu(n_{c \bar c} u^\mu) = \left[(\langle\sigma_{q{\bar
q}\rightarrow c\bar c}v\rangle + \langle\sigma_{q{\bar
q}\rightarrow c\bar c g}v\rangle)(n_{q}^{\rm eq})^2\right.\nonumber\\
&&+\left.\frac{1}{2}(\langle\sigma_{g g\rightarrow c\bar c}v\rangle
+ \langle\sigma_{g g\rightarrow c\bar c g}v\rangle )(n_{g}^{\rm
eq})^2\right
]\left[1-\left(\frac{n_c}{n_c^{\rm eq}}\right)^2\right].\nonumber\\
\end{eqnarray}

\subsection{collision dynamics at LHC}

In central heavy-ion collisions at RHIC, the particle distribution
at mid-rapidity is approximately uniform, so the Bjorken
boost-invariant hydrodynamic solution is a good approximation for
describing the time evolution of the hot dense matter formed in
these collisions \cite{Bjorken83}. Such a model indeed gives a good
description of the experimental data at RHIC \cite{Kolb03}. Also,
recent studies based on the AdS/CFT correspondence have demonstrated
that at late proper times the system produced in relativistic
heavy-ion collisions in the strongly-coupled regime described by
gauge-gravity duality exhibits an energy density scaling with
characteristics similar to the Bjorken hydrodynamic solution
\cite{AdS-1,AdS-2}. We thus expect that the longitudinal dynamics in
heavy ion collisions at LHC is also boost invariant. To include the
effect of transverse flow, we further include an accelerated
transverse expansion, resulting in a cylindrical symmetry in the
geometry of the collision. In terms of the cylindrical coordinates
$r$ and $\varphi$ as well as the proper time $\tau$ and the
space-time rapidity $\eta$ defined by \cite{Koch02,Ko3}
\begin{eqnarray}
\tau=\sqrt{t^2-z^2},\ \ \eta=\frac{1}{2}\ln\frac{t+z}{t-z},
\end{eqnarray}
we then have $u^{\eta}=u^{\varphi}=0$. If we further assume a
uniform density distribution in the transverse plane and take the
average over the radial coordinate, the left hand side of the rate
equation Eq.(\ref{rate-eq:2}) can then be expressed as:
\begin{eqnarray}
\frac{1}{\tau R^2(\tau)}\frac{\partial}{\partial\tau}(\tau
R^2(\tau)n_{c \bar c} \langle u^\tau\rangle),
\end{eqnarray}
where $R(\tau)$ denotes the transverse radius of the system, and
$\langle u^\tau\rangle$ is the averaged $\tau$ component of the four
velocity defined as
\begin{eqnarray}
\langle u^\tau \rangle = \frac{2}{R^2 (\tau)} \int_0^{R(\tau)} dr\,
r u^\tau(r)\,.
\end{eqnarray}

At mid-rapidity, the four velocity of a fluid element $u^\mu$ can be
described by two independent boosts in longitudinal and radial
directions \cite{Heinz93}, with the radial flow velocity $\beta_r$
given by
\begin{eqnarray}
u^\tau = \gamma_r = \frac{1}{\sqrt{1 -\beta_r^2}} \,.
\end{eqnarray}
Assuming the usual ansatz for the radial expansion velocity
\cite{Bondorf78,Ko3,Heinz93,Koch02}
\begin{eqnarray}
\beta_r (\tau,r) = \frac{dR}{d\tau}\left( \frac{r}{R} \right)\,,
\end{eqnarray}
we have
\begin{eqnarray}
\langle u^\tau \rangle = \int_0^1 dy\,\frac{1}{\sqrt{1-(dR/d\tau)^2
y}}.
\end{eqnarray}

With the time evolution of the transverse radius of the
fire-cylinder taken to be \cite{chen04}
\begin{eqnarray}
R(\tau) = R_0+ a(\tau-\tau_0)^2/2, \label{fireball-radius}
\end{eqnarray}
where $R_0$ and $\tau_0$ are the initial radius and proper time of
the fire-cylinder, respectively, and $a$ denotes the transverse
acceleration, the volume of the fireball at proper time $\tau$ is
then given by
\begin{eqnarray}
V(\tau)=\pi R^2(\tau)\tau  \,\,\, .\label{fireball-volume}
\end{eqnarray}
In the following, we choose the initial transverse radius of the
fire-cylinder to be the radius of the colliding Pb nucleus, i.e.,
$R_0=7.0~{\rm fm}$, and a transverse acceleration $a=0.1~c^2$/fm in
order to obtain a reasonable final transverse flow velocity.

\subsection{time evolution of the temperature of the quark-gluon
plasma}

To estimate the initial temperature of produced quark-gluon plasma,
we use two different models: A Multi-phase Transport (AMPT) model
\cite{ampt} and the model based on the Color Glass Condensate
\cite{lappi}. Using the AMPT model, we obtain an initial transverse
energy of about 3000 GeV per unit rapidity at the midrapidity in
central Pb+Pb collisions at $\sqrt{s_{NN}}=5.5~{\rm TeV}$. Most of
this initial energy is concentrated within a transverse radius of
about 4.7 fm that is smaller than the radius of the colliding nuclei
as a result of their surface diffuseness. If we take an earlier
initial proper time of $\tau_0=0.2~{\rm fm}/c$ than that at RHIC,
the initial energy density of produced quark-gluon plasma is then
about
\begin{equation}
\epsilon_0\approx \frac{dE_T/dy}{\pi R_0^2\tau_0}\approx
\frac{3000}{\pi\times 4.7^2\times 0.2}\approx 226~{\rm GeV/fm}^3.
\end{equation}
Using the relation
\begin{equation}
\epsilon=\frac{37\pi^2T^4}{30}\approx(T/160~{\rm MeV})^4~{\rm
GeV/fm}^3
\end{equation}
for the energy density of a quark-gluon plasma with massless quarks
and gluons, the initial temperature of the quark-gluon plasma formed
at LHC is thus $T_0\approx 620~{\rm MeV}$. This number turns out to
be very close to that predicted by the Color Glass Condensate model.
As shown in Ref.\cite{lappi}, the glasma formed after Pb+Pb
collisions at $\sqrt{s_{NN}}=5.5~{\rm TeV}$ has an energy density of
about $\epsilon\sim 700~{\rm GeV/fm^3}$ at a proper time $\tau\sim
0.07~{\rm fm}/c$. The glasma then evolves into a thermalized
quark-gluon plasma at a proper time $\tau_0$. Assuming that the
energy density decreases inversely with the proper time during this
stage, its value at $\tau_0\sim 0.2~{\rm fm}/c$ is then
$\epsilon_0\approx 245~{\rm GeV/fm}^3$, corresponding to an initial
temperature $T_0\approx 633~{\rm MeV}$. Since the result from the
Color Glass Condensate model depends on the fourth power of the
saturation momentum, the predicted initial temperature thus has a
large uncertainty. The same is true for the AMPT model as the
predicted total transverse energy is sensitive to the parton
structure function of the nucleon, the initial nuclear shadowing,
and the saturation of produced gluons. Therefore, we will consider
in the following also the scenario with an initial energy density
which is 50\% or a factor of two larger larger than above value,
corresponding to an initial temperature of about 700 or 750 MeV,
respectively.

We note for central Au+Au collisions at $\sqrt{s_{NN}}=200~{\rm
GeV}$ at RHIC, the total transverse energy from the AMPT model is
about 1000 GeV. Taking a proper time $\tau_0=0.5~{\rm fm}/c$ for the
formation of an equilibrated quark-gluon plasma as required by the
hydrodynamic model to explain the observed hadron elliptic flows
\cite{teaney,huovinen,hirano}, the initial energy density and
temperature of produced quark-gluon plasma are then
$\epsilon_0\approx 33~{\rm GeV/fm}^3$ and $T_0\approx 383~{\rm
MeV}$, respectively. These values are again similar to those from
the Color Glass Condensate model, which gives an energy density
$\epsilon\approx 130~{\rm GeV/fm}^3$ at $\tau_0=0.1~{\rm fm}/c$,
leading thus to $\epsilon_0\approx 26~{\rm GeV/fm}^3$ and
$T_0\approx 361~{\rm MeV}$ at $\tau_0=0.5~{\rm fm}/c$. With such a
low initial temperature, charm production from the quark-gluon
plasma is thus negligible according to the production rates shown in
Fig.~\ref{rate}.

\vspace{0.6cm}
\begin{figure}[ht]
\centerline{
\includegraphics[width=2.5in,height=2.5in,angle=0]{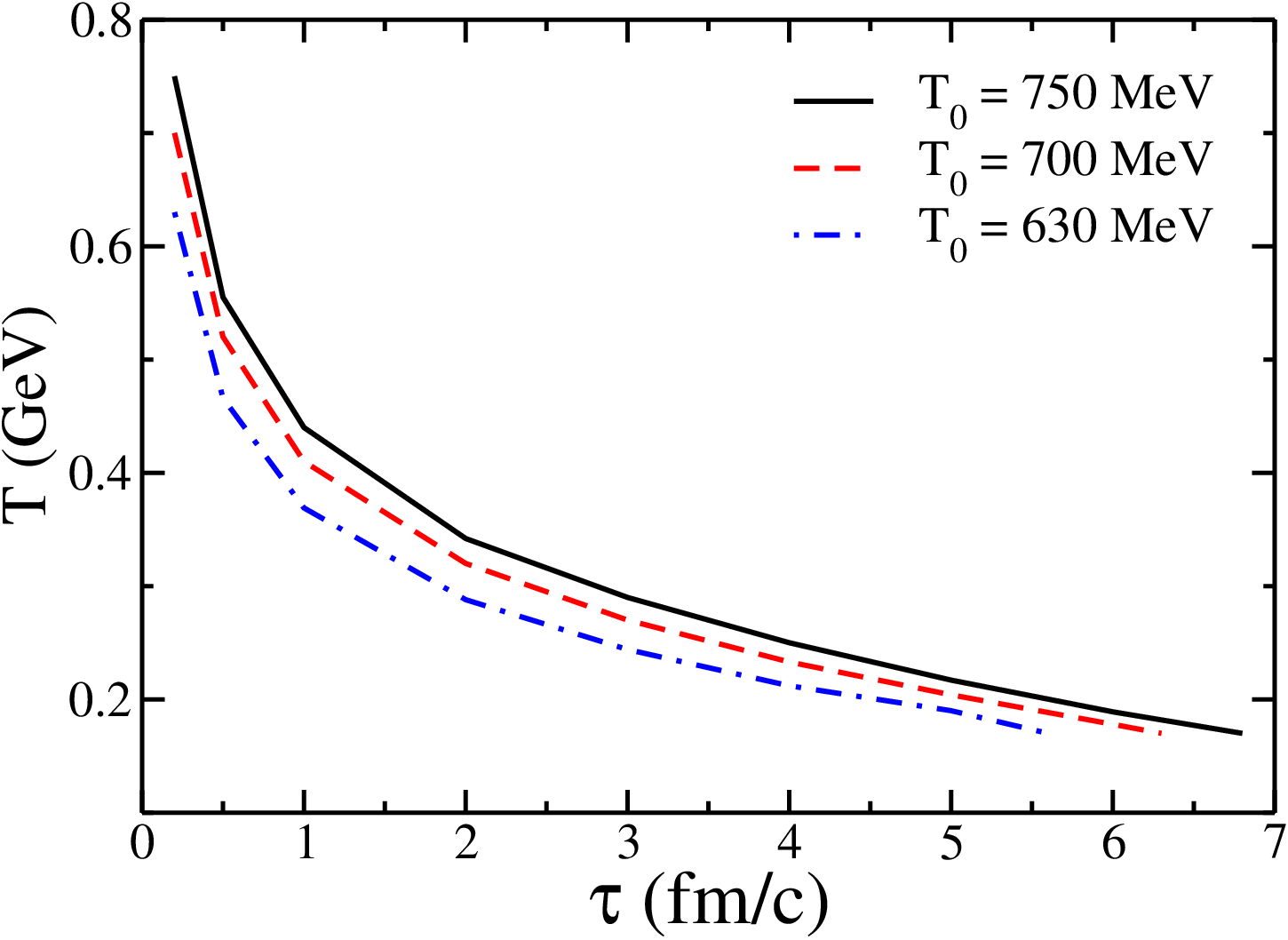}}
\caption{(Color online) Time evolution of the temperature of the
quark-gluon plasma formed in Pb+Pb collisions at
$\sqrt{s_{NN}}=5.5~{\rm TeV}$ for different initial temperatures and
same initial proper time $\tau_0=0.2~{\rm fm}/c$.}
\label{fireball-tem}
\end{figure}

For the time evolution of the temperature of produced quark-gluon
plasma at LHC, we determine it using the entropy conservation, and
the results are shown in Fig.~\ref{fireball-tem} for the three
initial temperatures $T_0 = 630$, $700$, and $750$ MeV. In all
cases, the critical temperature for the quark-gluon plasma phase
transition to the hadronic matter is taken to be $T_C = 170~{\rm
MeV}$. It is seen that the temperature decreases very quickly with
the proper time for all three initial temperatures, dropping below
400 MeV after about 0.5, 1, and 1.5 fm/$c$ for $T_0 = 630$, 700, and
750 MeV, respectively.

\section{Results}

For central Pb+Pb collisions at $\sqrt{s_{NN}}=5.5$ TeV at LHC, the
number of charm pairs produced in one unit rapidity from initial
hard nucleon-nucleon collisions is about 20 at midrapidity according
to the next-to-leading order pQCD calculations \cite{stat03}.
Neglecting the contribution from pre-thermal production, we solve
the rate equation with this initial number of charm quarks pairs and
certain initial temperature and proper time for the produced
quark-gluon plasma. We will vary the initial temperature and proper
time to study their effects on the final number of charm quark pairs
produced in these collisions.

\vspace{0.8cm}
\begin{figure}[ht]
\centerline{
\includegraphics[width=2.5in,height=2.5in,angle=0]{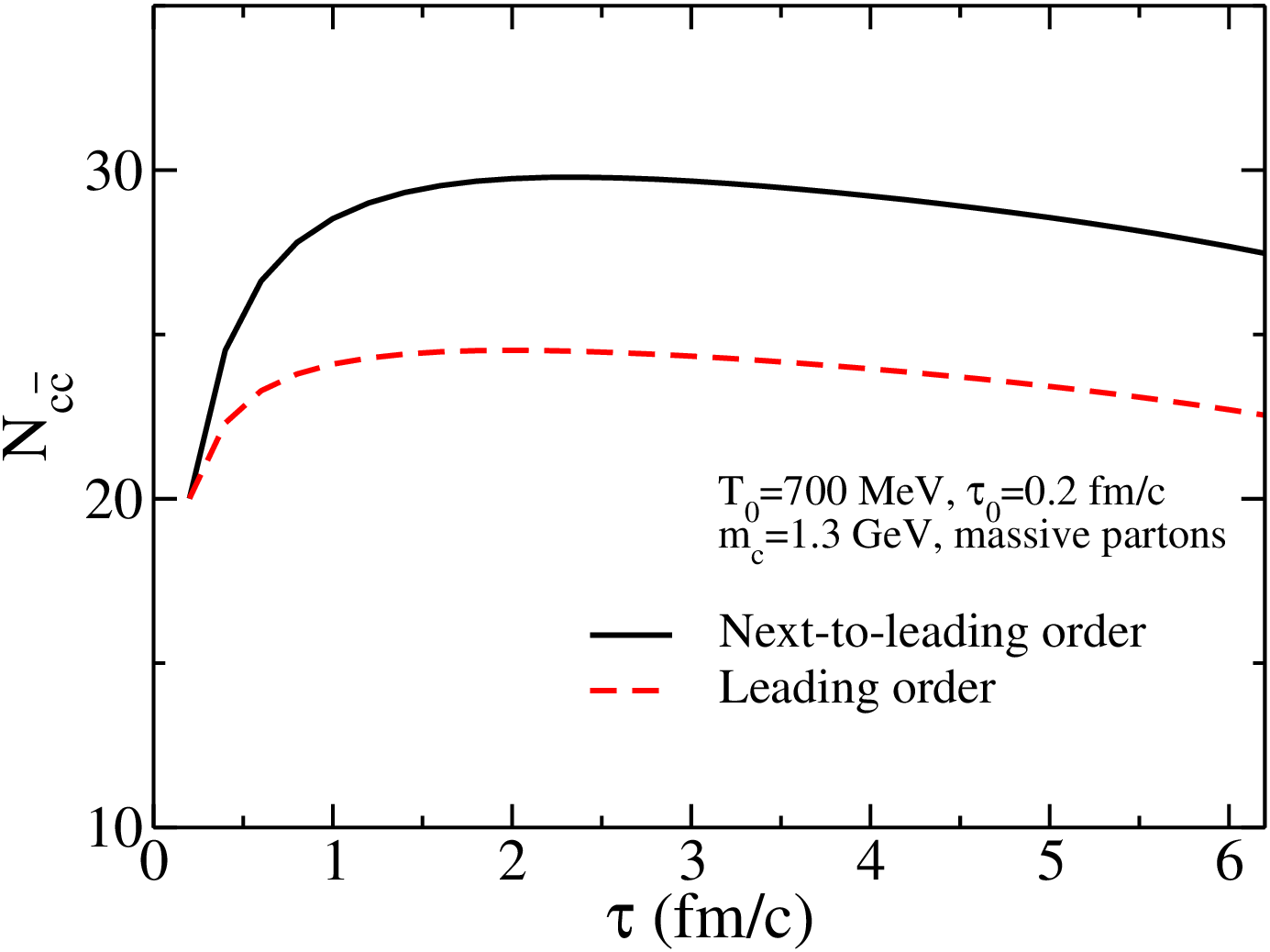}}
\caption{(Color online) Number of charm pairs as a function of
proper time in central Pb+Pb collisions at $\sqrt{s_{NN}}=5.5$ TeV
in the leading order (dashed line) and the next-to-leading order
(solid line) in QCD.} \label{number}
\end{figure}

We first consider the case of an initial temperature $T_0=700~{\rm
MeV}$ and proper time $\tau_0=0.2~{\rm fm}/c$. For the charm quark
mass, we take it to be $m_c=1.3$ GeV as that extracted from the
experimental data in $e^+e^-$ and $pp$ collisions. To take into
account the medium effect on quarks and gluons, we use the thermal
masses given in Eq.(\ref{thermal}). The time evolution of the total
number of charm quark pairs obtained from these initial conditions
in central Pb+Pb collisions at $\sqrt{s_{NN}}=5.5$ TeV is shown in
Fig.~\ref{number}. It is seen that including thermal production
enhances the total number of final charm pairs as compared to that
from initial direct production. The charm pair number reaches the
peak value at $\tau\sim 2$ fm/$c$ and then deceases with the proper
time. At the critical temperature $T_{\rm C}=170$ MeV when the
proper time is $\tau_{\rm C}\approx 6.4$ fm/$c$, the number of charm
pairs is about $27$ in the next-to-leading order (solid line), which
is about 30\% larger than the number due to direct production from
initial hard collisions. The final charm pair number is reduced to
about $22$ if only thermal production from leading-order
contribution is included as shown by the dashed line.

We note that the final charm quark pair number is much larger than
its chemically equilibrium value at $T_{\rm C}=170$ MeV, which is
about 4.7. This is not surprising as the time for charm pairs to
reach their equilibrium number $n_c^{\rm eq}$ at certain
temperature, given by $\tau_{\rm eq} =n_{c\bar c}^{\rm
eq}/2R_{c\bar{c}}$ with $R_{c\bar{c}}$ denoting the production rate
at that temperature, increases dramatically with decreasing
quark-gluon plasma temperature from a value of a few fm/$c$ at
$T=700$ MeV to a few thousands of fm/$c$ at $T_c=170$ MeV. It is
thus not possible for charm quarks to reach chemical equilibrium
during the finite lifetime of the produced quark-gluon plasma at
LHC. Since the initial number of charm quark pairs produced from
hard scattering of initial nucleons is large at LHC and the charm
production rate is also larger during the early stage of the
quark-gluon plasma when its temperature is high, there are more
charm pairs in the quark-gluon plasma than the equilibrium number at
the critical temperature. As the quark-gluon plasma expands and
cools, the charm annihilation rate decreases, making it less likely
to destroy the produced charm pairs and leading thus to an
over-saturation of the charm abundance at the critical temperature.

\vspace{0.6cm}
\begin{figure}[ht]
\centerline{
\includegraphics[width=2.5in,height=2.5in,angle=0]{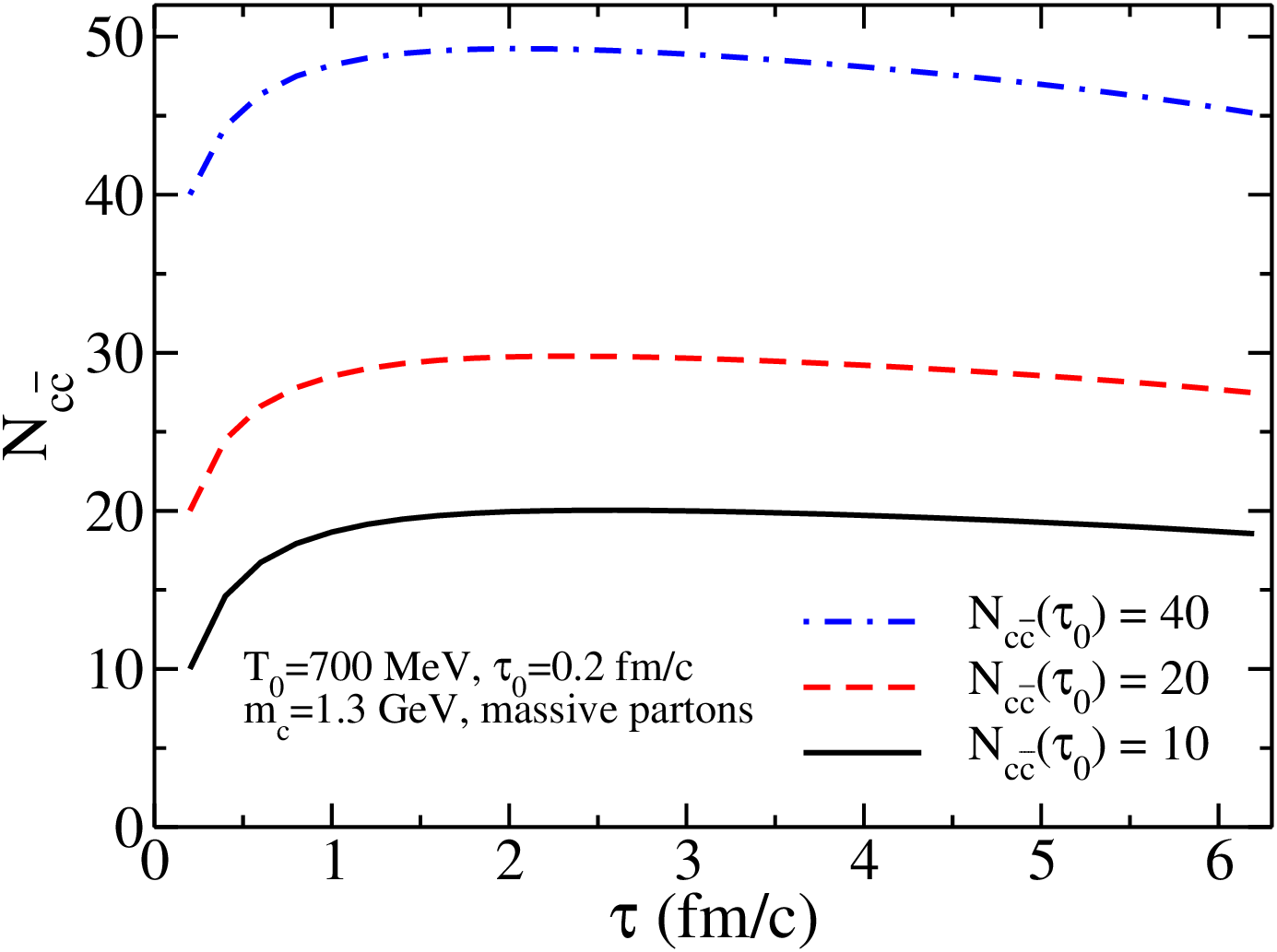}}
\caption{(Color online) Number of charm pairs as a function of
proper time in central Pb+Pb collisions at $\sqrt{s_{NN}}=5.5$ TeV
for different number of charm quark pairs produced from initial hard
scattering.} \label{initial}
\end{figure}

Since there is substantial uncertainty in the charm quark pairs
produced from initial direct production \cite{stat03,Vogt95}, we
have also studied thermal charm production using different initial
charm quark pair numbers. Varying the initial charm pair number by a
factor of two, we have repeated above calculations, and the results
based on the next-to-leading order in QCD are shown in
Fig.~\ref{initial}. For initial numbers of charm quark pairs of 10,
20 and 40, the final numbers are found to be about 19, 27, and 45,
respectively. Thermal production of charm quarks from the
quark-gluon plasma thus becomes more important as the initial charm
pair number becomes smaller.

Returning to the case of 20 initial charm quark pairs, we note that
in above calculations, we have used massive quarks and gluons given
by the thermal QCD calculations. To see how these masses affect
thermal charm production, we have carried out similar calculations
as above with massless quarks and gluons but same initial conditions
of $T_0=700~{\rm MeV}$ and $\tau_0=0.2~{\rm fm}/c$ as well as the
charm quark mass of $m_c=1.3~{\rm GeV}$. We find that the resulting
number of charm quark pairs at $T_C=170~{\rm MeV}$ differs hardly
from the massive case. The reason for this is that although using
massless quarks and gluons reduces the thermal averaged charm
production cross sections, the effect is largely compensated by the
larger quark and gluon densities if they are massless.

\vspace{0.6cm}
\begin{figure}[ht]
\centerline{
\includegraphics[width=2.5in,height=2.5in,angle=0]{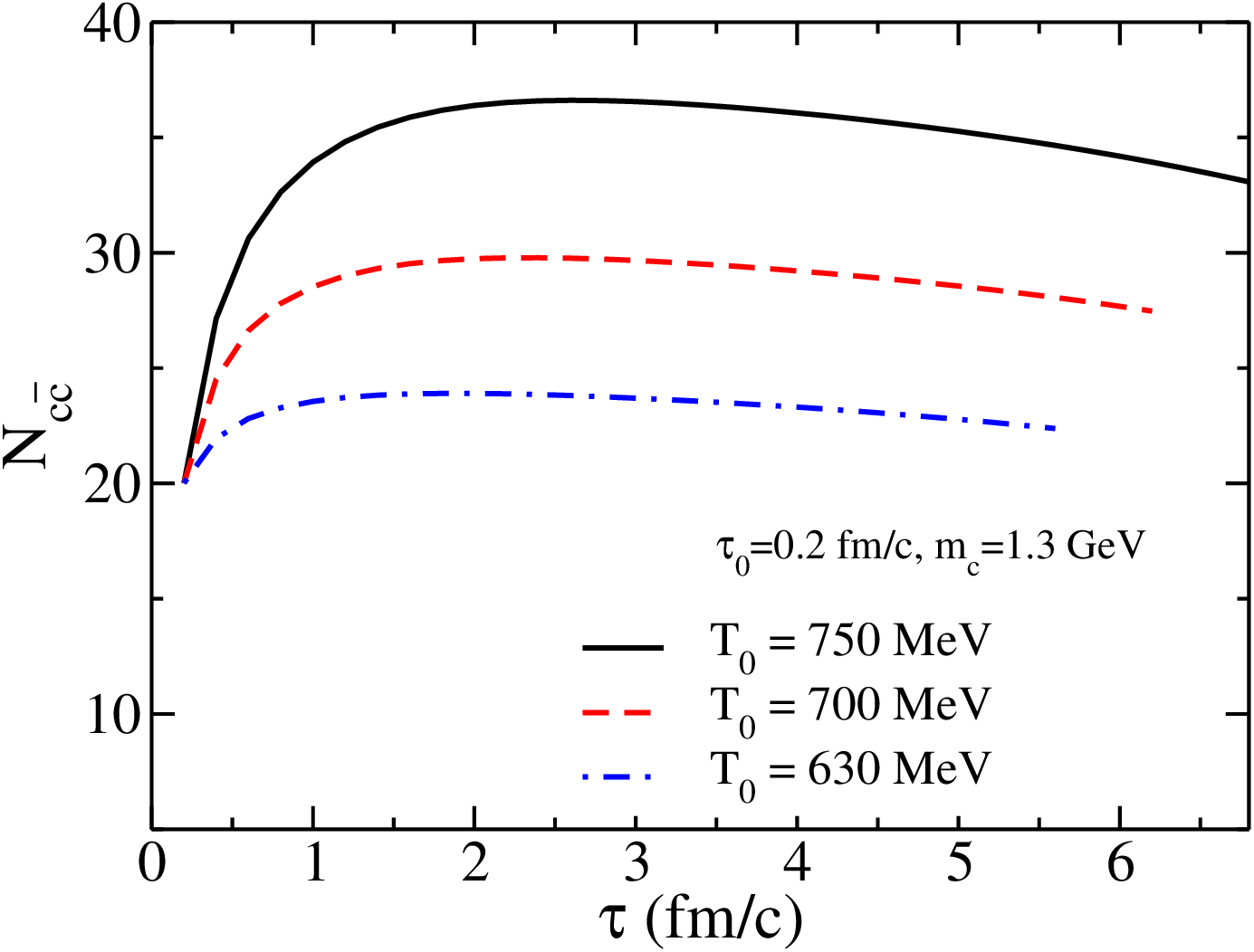}}
\caption{(Color online) Total number of charm pairs as a function of
proper time in central Pb+Pb collisions at $\sqrt{s_{NN}}=5.5$ TeV
for different initial temperatures of the quark-gluon plasma but
same initial proper time $\tau_0=0.2~{\rm fm/}c$.}
\label{temperature}
\end{figure}

As previously mentioned, the charm production rate exhibits an
exponential increase with the temperature of the quark-gluon plasma.
The total number of charm quark pairs produced in heavy ion
collisions thus depends on the initial temperature of the expanding
quark-gluon plasma. In Fig.~\ref{temperature}, we show the total
number of charm pairs as a function of proper time in central Pb+Pb
collisions at $\sqrt{s_{NN}}=5.5$ TeV for different initial
temperatures of the quark-gluon plasma but same initial proper time
$\tau_0=0.2~{\rm fm}/c$. It is seen that the final number of charm
quark pairs at the critical temperature $T_C=170$ MeV calculated in
the next-to-leading-order decreases to about 22 for $T_0=630~{\rm
MeV}$ and increases to about 33 for $T_0=750~{\rm MeV}$.

We have also considered the effect of initial proper time on thermal
charm production by using $\tau_0=0.5~{\rm fm/}c$ as that in heavy
ion collisions at RHIC. The predicted initial temperature from the
Color Glass Condensate model is then $T_0=500~{\rm MeV}$. Results
for this initial temperature as well as those for the temperatures
of 560, and 600 MeV, which correspond, respectively, to an increase
of the initial energy by 50\% and 100\% of that for $T_0=500$ MeV,
show that the final total number of charm quark pairs is only
slightly reduced compared to that for an initial proper time
$\tau_0=0.2~{\rm fm}/c$, i.e., 21, 26, and 32 for above three
temperatures.

The charm quark mass also affects the contribution of thermal charm
production. With $m_c=1.5~{\rm GeV}$ and the initial proper time
$\tau_0=0.2~{\rm fm}/c$, the final total number of charm quark pairs
is reduced to about 20, 22 and 25 for the three initial temperatures
$T_0=630$, 700, and 750 MeV, respectively.

\section{Summary and Discussions}

In this paper, we have carried out the first calculation of thermal
charm production at next-to-leading-order in QCD. Modeling central
heavy ion collisions at LHC by a schematic longitudinally boost
invariant and transversely expanding fire-cylinder of quark-gluon
plasma, we have evaluated the number of charm quark pairs produced
in these collisions. With an initial temperature of 700 MeV for an
equilibrated quark-gluon plasma at an initial proper time of 0.2
fm/$c$ and a charm quark mass of 1.3 GeV, we have obtained about
30\% enhancement in the production of charm quarks than that
produced directly from initial hard collisions, if the latter is
taken to be 20 pairs at midrapidity according to the next-to-leading
order pQCD calculations. About equal contributions are obtained from
the leading order and the next-leading order processes. This result
is, however, sensitive to the initial conditions for the produced
quark-gluon plasma as well as the charm quark mass. The enhancement
is increased to about 80\% if the initial temperature is increased
to 750 MeV but reduced to about 10\% if the initial temperature is
decreased to 630 MeV. Delaying the proper time at which a
thermalized quark-gluon plasma is formed does not affect, however,
much thermal charm quark production as the effect due to decreased
initial temperature is compensated by that from the increased volume
of the quark-gluon plasma. Changing the charm quark mass has, on the
other hand, a large effect on thermal charm quark production from
the quark-gluon plasma. With a larger charm quark mass of
$m_c=1.5~{\rm GeV}$, thermal charm quark production from the
quark-gluon plasma becomes unimportant even for an initial
temperature of $T_0=700~{\rm MeV}$. Finally, thermal charm
production from the quark-gluon plasma becomes more important if the
number of directly produced charm pairs from initial hard scattering
is small. These results are not only of interest in their own right
but also useful for understanding charmonium production in
relativistic heavy ion collisions as its production from the
quark-gluon plasma is proportional to the square of the charm quark
numbers.

\begin{acknowledgments}
This work was supported in part by the US National Science
Foundation under Grant No. PHY-0457265 and the Welch Foundation
under Grant No. A-1358. Ben-wei Zhang was further supported by the
National Natural Science Foundation of China under project No.
10405011 and by MOE of China under project IRT0624.
\end{acknowledgments}


\end{document}